\documentclass[10pt]{iopart}
\usepackage{iopams}
\usepackage{graphicx}
\usepackage{dcolumn}
\usepackage{bm}
\usepackage[colorlinks = true, allcolors = blue]{hyperref}
\usepackage{verbatim}
\usepackage{soul}
\usepackage{float}
\usepackage{xcolor}
\usepackage{multirow}

\begin{document}
\title{Dispersive interactions between standard and Dirac materials and the role of dimensionality}

\author{Dai-Nam Le$^{1,2,3}$, Pablo Rodriguez-Lopez$^4$, Lilia M. Woods$^{1,5}$}

\address{$^1$Department of Physics, University of South Florida, Tampa, Florida 33620, USA}
\address{$^2$Atomic~Molecular~and~Optical~Physics~Research~Group, Advanced Institute of Materials Science, Ton Duc Thang University, Ho~Chi~Minh~City 700000, Vietnam}
\address{$^3$Faculty of Applied Sciences, Ton Duc Thang University, Ho~Chi~Minh~City 700000,~Vietnam}
\address{$^4${\'A}rea de Electromagnetismo and Grupo Interdisciplinar de Sistemas Complejos (GISC), Universidad Rey Juan Carlos, 28933, M{\'o}stoles, Madrid, Spain}
\address{$^5$Author to whom all correspondence should be addressed}

\ead{ledainam@tdtu.edu.vn (Dai-Nam Le)}
\ead{pablo.ropez@urjc.es (Pablo Rodriguez-Lopez)}
\ead{lmwoods@usf.edu (Lilia M. Woods)}

\begin{abstract}
The van der Waals interaction plays a prominent role between neutral objects at separations where short ranged chemical forces are negligible. This type of dispersive coupling is determined by the interplay between geometry and response properties of the materials making up the objects. Here, we investigate the van der Waals interaction between 1D, 2D, and 3D standard and Dirac materials within the Random Phase Approximation, which takes into account collective excitations originating from the electronic Coulomb potential. A comprehensive understanding of characteristic functionalities and scaling laws are obtained for systems with parabolic energy dispersion (standard materials) and crossing linear bands (Dirac materials). By comparing the quantum mechanical and thermal limits the onset of thermal fluctuations in the van der Waals interaction is discussed showing that thermal effects are significantly pronounced at smaller scales in reduced dimensions. 

\end{abstract}

\submitto{Journal of Physics: Materials}
\maketitle
\ioptwocol

\section{Introduction}

The exchange of electromagnetic excitations of two objects brought together results in a dispersive ubiquitous force whose sign and magnitude depend on the materials properties and geometry \cite{Woods2016, Klimchitskaya2009}. At short separations, this exchange is instantaneous and the interaction is termed as the van der Waals (vdW) force whose main contribution comes from exponentially decaying surface modes. The vdW interaction is important for the stability of layered materials and composites made of chemically inert components. It also plays a significant role in the organization and complex behavior of biological systems, including lipids, membranes, and proteins. 

In the simplest possible way, pair-wise summation of interatomic Lennard-Jones type of potentials may be used to estimate the vdW interaction between molecules and/or between extended objects \cite{Girifalco2000, Ulbricht2002}. However, such a simplified approach does not take into account collective effects, which become significant for nondilute systems. By placing objects close to each other, their density fluctuations become affected by the mutual Coulomb interaction, which can result in different behavior of the vdW force when compared with results obtained by the pairwise summation method \cite{Dobson2014}.  To capture electron correlation effects in the response properties of the interacting materials as well as their vdW force one can use the Random Phase Approximation (RPA), which is a second-order perturbative nonretarded approach with respect to the Coulomb interaction. This method has been successfully applied to a variety of systems  \cite{Sernelius1998, Dobson2006, Dobson2009, Santos2009, Woods2014}. Advanced computational techniques to calculate vdW interactions from first principles based on the RPA approach as well as other techniques have also been developed and shown to give accurate results in many weakly bound systems  \cite{Dobson2012, Tkatchenko2014,Ambrosetti2016,Ambrosetti2017,Ambrosetti2018a,Ambrosetti2018b}.

Understanding how distinct properties of the materials affect the vdW interaction in different dimensions is particularly important in the context of many recent discoveries of topologically nontrivial systems \cite{Liu2019, Felser2021}. Given that the electronic structure and dimensionality determine the polarization properties of materials, it is important to investigate this interplay in the ubiquitous vdW force. Distinguishing between materials with parabolic and linear energy dispersions in one, two, and three dimensions can provide useful comprehensive insight in various asymptotic regimes of dispersive interactions involving standard and Dirac systems. Results from such investigations can also serve as a guide to interpret first principles simulations and experimental meausrements for dispersive interactions for various systems.

In this paper, we utilize the RPA approach to calculate the density response function for 1D, 2D, and 3D materials with parabolic and linear energy dispersions. The vdW interaction is obtained based on the density-density correlator by treating the mutual Coulomb coupling as a perturbation \cite{Walecka}. By analyzing the characteristic behaviors in the dielectric response, we are able to provide a comprehensive description of the vdW interaction by elucidating the roles of electronic structure and dimensionality. By comparing the obtained results in the quantum mechanical and thermal limits, the onset of thermal fluctuations in the different dimensionalities is discussed.  Unique features in terms of scaling laws and their dependence upon the Fermi level in the various dimensions are also obtained for standard and Dirac materials.

\section{Dispersive Interactions - the RPA limit}

\begin{figure}[ht]
    \centering
    \includegraphics[width = 0.45 \textwidth]{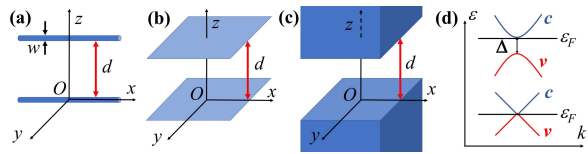}
    \caption{Schematics of two identical isotropic (a) $1$D, (b) $2$D, and (c) $3$D materials in vacuum separated by a distance $d$ along the $z$-axis. The $1$D systems are taken as wires with a diameter $w \ll d$. The $2$D layers are infinitely thin. (d) Schematics of the conduction ($c$) and valence ($v$) parabollic and crossing linear bands. A band gap $\Delta$ separates the $c$ and $v$ regions. The Fermi level is denoted as $\varepsilon_F$.}
    \label{fig:1}
\end{figure}

The interacting objects in 1D, 2D, and 3D geometries are schematically shown in Fig. \ref{fig:1}. In each case, the objects are placed in vacuum and are separated by a distance $d$ along the $z$-axis. The wires in Fig. \ref{fig:1}a are with diameters such that $w \ll d$ and the layers in Fig. \ref{fig:1}b are assumed to be infinitely thin. The 3D objects in Fig. \ref{fig:1}c are taken to be semi-infinite.

At smaller separations, the vdW dispersive force is determined by the instantaneous exchange by exponentially decaying modes localized on the surface of each object. These are $p$-polarized electromagnetic excitations and the interaction energy can be found as a result of the Coulomb potential between the density fluctuations  in the objects \cite{Walecka, Sernelius1998},
 \begin{eqnarray}\label{eqn:general}
     F^{(N)}  = && \int \frac{d^N \mathbf{q}}{(2\pi)^N} \times \nonumber\\
    && \times \left. \frac{\partial \mathcal{V}^{(N)} \left(\mathbf{q},  z\right)}{\partial z} \right|_{z=d} \left\langle n_1 \left( \mathbf{q} \right) n_2 \left( - \mathbf{q} \right) \right\rangle ^{(N)}.
 \end{eqnarray}
In the above integral over the wave vector ${\bf q}$, $N$ denotes the dimensionality and $\mathcal{V}^{(N)}$ is the Coulomb interaction between the objects. Also, $\left\langle n_1 \left( \mathbf{q} \right) n_2 \left( - \mathbf{q} \right) \right\rangle ^{(N)}$ is the density-density correlator determined by the density fluctuations $n_1(\bf q)$ and $n_2(-\bf q)$ in the two objects (denoted as (1), (2)). The density-density correlator at finite temperature $T$ in the RPA takes into account the collective screening effects and it is taken at imaginary Matsubara frequencies  $\omega _l =  2 \pi k_B T l / \hbar$  for the calculations of the vdW interaction  \cite{Walecka,Sernelius1998}, 
\begin{eqnarray}\label{eqn:density}
   && \left\langle n_1 \left( \mathbf{q} \right) n_2 \left( - \mathbf{q} \right) \right\rangle ^{(N)} = \nonumber\\
   && \quad \quad =  k_B T \sum \limits_{l = 0}^{+ \infty}{}^{\displaystyle\prime} \chi _{1}^{(N)} (\mathbf{q}, \imath \omega _l) \mathcal{W} (\mathbf{q}, d) \chi _{2}^{(N)} (\mathbf{q}, \imath \omega _l), \nonumber\\ 
  && \quad \quad =  \frac{2 k_B T}{\mathcal{V}^{(N)} \left( \mathbf{q},d \right)} {\sum \limits_{l = 0}^{+ \infty } } {}^{\displaystyle\prime} \left\{ \left[ \frac{v_0^{(N)} \left( \mathbf{q} \right)}{\mathcal{V}^{(N)} \left(\mathbf{q}, d \right)}\right]^2 \times \right. \nonumber\\
  && \quad \quad \quad \left. \times \left[1 - \frac{1}{v_0^{(N)} \left( \mathbf{q} \right) \chi^{(N)}_{0} \left(\mathbf{q}, \imath \omega _l \right)} \right]^2  - 1 \right\}^{-1},
\end{eqnarray}
where $v_{0}^{(N)} \left( \mathbf{q} \right)$ is the Coulomb interaction of electrons inside each material in reciprocal $\mathbf{q}$ space and $\chi_{0}^{(N)} \left(\mathbf{q}, \omega \right)$ is the bare polarization function. Within the RPA method, the screened Coulomb interaction $\mathcal{W}_{0}^{(N)} \left( \mathbf{q} , z \right) = \frac{\mathcal{V}_{0}^{(N)} \left( \mathbf{q}, z \right)}{1 - \chi _{1}^{(N)} (\mathbf{q}, \imath \omega _l) \mathcal{V}_{0}^{(N)} \left( \mathbf{q}, z \right) \chi _{2}^{(N)} (\mathbf{q}, \imath \omega _l)}$ and the screened polarization function  $\chi_{1,2}^{(N)} \left(\mathbf{q},i \omega_l \right) = \frac{\chi_{0}^{(N)} \left(\mathbf{q}, i\omega_l \right)}{1 - v_{0}^{(N)} \left( \mathbf{q} \right) \chi_{0}^{(N)}( \left(\mathbf{q}, i\omega_l \right)}$ 
are obtained by summing connected bubble diagrams within the Feynman diagrammatic representation \cite{Walecka,Hwang2007,Sarma2009}.  The prime in the summation indicates that the $l=0$ term is multiplied by $1/2$.

It is apparent that for the interaction calculations, we need to distinguish the Coulomb coupling in each object and the Coulomb coupling between the objects. These can be derived from  standard electrostatics following Gauss law (see Supplementary Information) and the results are summarized in Table \ref{tab:1} for the 1D, 2D and 3D cases from Fig. \ref{fig:1}.

\begin{table*}
    \centering
    \caption{Coulomb coupling inside each object, $v_{0}^{(N)} (\mathbf{q})$, and between the objects, $\mathcal{V}^{(N)} (\mathbf{q},z)$, where $e$ is the electron charge, 
$\gamma_E$ - Euler-Mascheroni number, $w$ - width of the 1D object, $q_\perp=\sqrt{q_x^2+q_y^2}$, and $K_0(y)$ - the modified Bessel function of second kind. }
    \label{tab:1}
\resizebox{1 \textwidth}{!}{
    \begin{tabular}{cccc}
    \hline\hline
    & $1$D & $2$D & $3$D \\
    \hline\hline
   $v_{0}^{(N)} (\mathbf{q})$ & $v^{(1D)}_{0} (q_x) = 2 e^2 \left[ \ln \left( \frac{2}{|q_x| w} \right) - \gamma _E \right] $ & $v^{(2D)}_{0} (\mathbf{q}) = \frac{2 \pi e^2}{q_\perp} $   &  $v^{(3D)}_{0} (\mathbf{q}) = \frac{4 \pi e^2}{q^2}  $ \\
    \hline
    $\mathcal{V}^{(N)} (\mathbf{q},z)$ & $\mathcal{V}^{(1D)} (q_x,z) = 2 e^2 K_0 \left( |q_x| z \right)$ & $\mathcal{V}^{(2D)} (\mathbf{q},z) = \frac{2 \pi e^2}{ q_\perp} e^{- q_\perp z}$ & $\mathcal{V}^{(3D)} (\mathbf{q},z) = \frac{4 \pi e^2}{q^2} e^{- q_{\perp} z} $\\
    \hline\hline
    \end{tabular}
}
\end{table*}

The polarization function $\chi_{0}^{(N)}\left(\mathbf{q}, \imath \omega \right) $ corresponds to a single bubble diagram and it is given as \cite{Walecka,Hwang2007,Sarma2009} 
\begin{eqnarray}\label{eqn:chi}
      \chi_{0}^{(N)} \left(\mathbf{q}, \imath \omega \right)  = g \sum\limits_{\mu, \nu} && \int  \frac{d^N \mathbf{k}}{(2\pi)^N} \mathcal{F}_{\nu, \mu} (\mathbf{k}, \mathbf{q}) \times \nonumber\\
      && \times \frac{f(\varepsilon _{\mu} (\mathbf{k}) ) - f( \varepsilon _{\nu} (\mathbf{k}+\mathbf{q}) )}{\imath \hbar \omega - \left[ \varepsilon _{\nu} (\mathbf{k}+\mathbf{q}) - \varepsilon _{\mu} (\mathbf{k})  \right] } ,
\end{eqnarray}
where $g$ is the degeneracy of the electronic states, $\varepsilon _{\mu,\nu}$ are the eigenenergies of the underlying Hamiltonian of the specific material making up the objects, and $f(\varepsilon_{\mu,\nu})$ are the Fermi distribution functions. The factor  $\mathcal{F}_{\nu, \mu} (\mathbf{k}, \mathbf{q}) = \left| \langle \nu, \mathbf{k} + \mathbf{q} | e^{\imath \mathbf{q} \cdot \mathbf{r}} |  \mu, \mathbf{k} \rangle \right|^2 = \left| \langle u_{\nu, \mathbf{k} + \mathbf{q}} | u_{\mu, \mathbf{k}} \rangle \right|^2$ corresponds to the overlap integral between the Bloch eigenstates $\left|u_{\mu, \mathbf{k}} \right\rangle$ of the Hamiltonian. 

Here we are interested in two types of systems: standard materials whose conduction ($c$) and valence ($v$) energy bands have a typical parabolic dispersion and Dirac materials characterized by crossing energy bands with a linear dispersion (schematics shown in Fig. \ref{fig:1}d). For the considered systems, the energy dispersion and overlap integrals can be calculated in the long-wave length limit \cite{Hwang2007, Sarma2009, Sachdeva2015, Ehrenreich1959} and the results are presented in Table \ref{tab:2}. Note that the range of validity for the linear dispersion $\varepsilon_{c,v}=\pm \hbar v_F k$ is taken to be determined by a bandwidth $\varepsilon_{max}$. For standard materials, the parabolic conduction and valence bands are separated by a band gap $\Delta$. One notes that when there is no band gap and the conduction and valence bands touch at a point at the Fermi level, the overlap integral is $\mathcal{F}_{\nu, \mu}=\delta_{\nu \mu}$.   For Dirac materials, the overlap integral depends on the angle between the ${\bf k}, {\bf q}$ wave vectors, $\theta_{{\bf k},{\bf q}}$, as well as the ratio $q^2/k^2$. Relevant details are given in the Supplementary Information.

\begin{table*}
    \centering
    \caption{Energy dispersion $\varepsilon _{c,v} (\mathbf{k})$ for the conduction ($c$) and valence ($v$) bands and corresponding overlap integral $\mathcal{F}_{\nu, \mu} (\mathbf{k},\mathbf{q})$ for standard and Dirac materials in the long-wavelength approximation ($\nu, \mu = (c,v)$). Here, $\Delta$ is the gap between the conduction and valence bands, $v_F$ is the Fermi velocity, and $\theta_{{\bf k},{\bf q}}$ is the angle between the ${\bf k}$ and ${\bf q}$ wave vectors.}
    \label{tab:2}
\resizebox{1 \textwidth}{!}{
    \begin{tabular}{ccc}
    \hline\hline
             & Energy dispersion & Overlap integral  \\
    \hline\hline
    Standard & $\varepsilon_c (\mathbf{k}) = \frac{\hbar ^2}{2 m} k^2$;  $\varepsilon_v (\mathbf{k}) = \Delta - \frac{\hbar ^2}{2 m} k^2$ & $\mathcal{F}_{\nu, \mu} (\mathbf{k},\mathbf{q}) = \delta _{\nu \mu} + \left( 1 - \delta _{\nu \mu} \right) \frac{\hbar^2 q^2}{2 m} \frac{\Delta}{ \left(2 \varepsilon_c (\mathbf{k}) + \Delta \right)^2 } $  \\
     \hline
    Dirac & $\varepsilon _c (\mathbf{k}) = -\varepsilon _v (\mathbf{k}) = \hbar v_F k$ &  $\mathcal{F}_{\nu, \mu} (\mathbf{k},\mathbf{q}) = \delta _{\nu \mu} + \left( 1 - \delta _{\nu \mu} \right) \frac{\hbar ^2 v_F^2 q^2}{4 \varepsilon_c^2 (\mathbf{k})} \sin^2 \theta _{\mathbf{k},\mathbf{q}} $ \\
    \hline\hline
    \end{tabular}
}
\end{table*}

\begin{figure}[ht]
    \centering
    \includegraphics[width = 0.45 \textwidth]{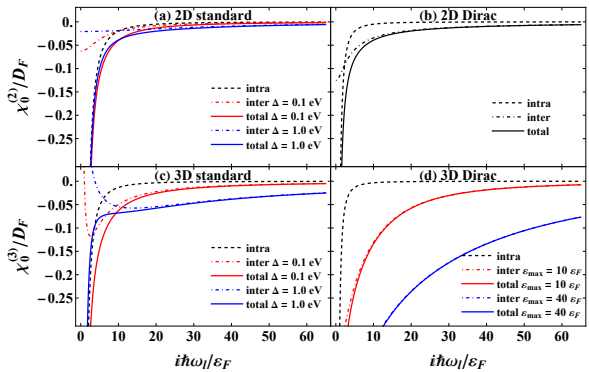}
    \caption{\label{fig:2}The intraband $\chi^{(N)}_{0,intra}$, interband $\chi^{(N)}_{0,inter}$, and total $\chi^{(N)}_{0}=\chi^{(N)}_{0,intra}+\chi^{(N)}_{0,intra}$ bare polarization versus imaginary frequency {at $q = k_F$} are shown for various 2D, 3D standard materials (a,c) and for 2D, 3D gapless Dirac materials (b,d). For all Dirac materials $\varepsilon _F = 50 \; \rm{ meV}$, $g = 4$, $v_F = c/300$, while $g = 2$, $m = m_e$ for standard materials. The bare polarization function is scaled by the density of states at the Fermi level according to: $D_F^{(2),D} = \frac{g}{2\pi} \frac{\varepsilon_F}{\hbar^2 v_F^2}$, $D_F^{(3),D} = \frac{g}{2\pi ^2} \frac{\varepsilon_F^2}{\hbar^3 v_F^3}$ for Dirac materials,  $D_F^{(2),S} = \frac{g}{4\pi} \frac{2m}{\hbar^2}$ and $D_F^{(3),S} = \frac{g}{4\pi^2} (\frac{2m}{\hbar^2})^{3/2} \varepsilon_F^{1/2}$ for standard materials.}
\end{figure}

Eq. \Eref{eqn:chi} captures interband and intraband transitions at imaginary frequency and the expressions from Table \ref{tab:2} allow explicit calculations for the bare polarization function for both types of materials in all dimensions. In addition to the general formulas for all materials and both types of transitions, analytical expressions for $\chi_{0}^{(N)}$ in the long wavelength approximation are shown in the Supplementary Information. Furthermore, the dependence of $\chi_{0}^{(N)}$ as a function of Matsubara frequencies scaled by the Fermi level is shown in Fig. \ref{fig:2} for some representative cases. We find that in standard materials, at small $\hbar\omega_l/\varepsilon_F$ the response function is dominated by intraband transitions, as expected. However, while in 2D the intraband contribution is almost a constant (essentially detemined by the gap), in 3D the intraband part changes sign at very small $\hbar\omega_l/\varepsilon_F$. Our calculations also show (see Supplementary Information), that in 2D massless Dirac material a cancellation between the intra and interband contributions occurs at $\varepsilon _F \ll \hbar \omega $; as a result, doping is a dominant factor in the small imaginary frequency range. 

In 3D Dirac materials, we find that the bandwidth specifying the validity of the linear band approximation for the energy dispersion, plays a prominent role. In fact, the interband contribution to $\chi_{0}$ depends explicitly on $\varepsilon_{max}$. This is consistent with previous results for the interband components of the optical conductivity tensor of $3$D Weyl semimetals \cite{Rodriguez2020}. It is also important to note that in both types of 1D materials, the response is dominated by the intraband part as we have found that $\chi^{(1)}_{0,inter} \left(\mathbf{q}, \imath \omega \right) \ll \chi^{(1)}_{0,intra} \left(\mathbf{q}, \imath \omega \right)$ (see Supplementary Information). This implies that metallic-like response behavior is expected for 1D systems. 

At this stage, we examine the quantum mechanical limit of the vdW interaction from Eq. \Eref{eqn:general} by substituting $\frac{2\pi k_B T}{\hbar}\sum\limits_{l=0}^{+\infty}{}^{\displaystyle\prime}$ with integration over frequency $\int\limits_0^{\infty} d\omega$ and taking the polarization function in Eq. \Eref{eqn:chi} at $T=0$. Asymptotic expressions for the vdW force can be found in the long wave approximation by also taking into account the Coulomb interactions in Table \ref{tab:1}, the energy band structures in Table \ref{tab:2} for the considered materials, and their polarization functions in the Supplementary Information. The obtained expressions are organized in Table  \ref{tab:3}. We find that the force for both types of 1D systems for which $\varepsilon_F>0$ is quite similar. In fact, the only difference is that  $\hbar v_F$ for Dirac materials is replaced by $2\hbar^2\varepsilon_F/m$ for standard materials, but the characteristic $[d \sqrt{\ln(2.2 d/w)}]^{-3}$ scaling law is the same. The obtained distance dependence is consistent with previous results for different types of 1D materials \cite{Dobson2006, Drummond2007, Woods2014}. This indicates that as long as $\varepsilon _F >0$ the particular electronic structure plays a secondary effect in the interaction.  For 1D parabolic materials with a finite gap, the interaction has a $d^{-6}$ dependence indicating London type of behavior, which was also found for graphene nanoribbons \cite{Stedman2014}. The coupling in this case can further be tuned by changing the band gap since $F^{(1) S}\sim \Delta ^{-2}$. We finally note that for 1D Dirac materials with $\varepsilon_F=0$, $\chi_0^{(1)} \to 0$ in long wave-length limit, thus the vdW force is negligible. 

\begin{table*}
    \centering
    \caption{Asymptotic scaling laws of the quantum mechanical and thermal vdW force of two identical Dirac and standard materials (degeneracy $g$) in 1D, 2D, and 3D. The dimensionless constants (analytical expressions given in the Supplementary Information) obtained from the calculations:  $ \mathcal{A}_{1S} \approx 0.01806$, $\mathcal{A}_{2S} \approx 0.01253$, $\mathcal{A}_{2D} \approx 0.007641$, $\mathcal{A}_{3S}\approx 0.438662$ and $\mathcal{A}_{3S}^{\prime}\approx 2.12266$. For 3D Dirac materials, the numerical constant $\mathcal{A}_{3D}$ depends on their Fermi velocity through the constant $\frac{g e^2}{\hbar v_F}$. Ex: $v_F = \frac{c}{300}$, $\mathcal{A}_{3D} \approx 0.397665$. {The expressions for the thermal vdW force are the same for both types of materials.} 
}   
 \label{tab:3}
\resizebox{1 \textwidth}{!}{    
    \begin{tabular}{llll}
    \hline\hline
       &  \multicolumn{1}{c}{Dirac materials} &  \multicolumn{1}{c}{Standard materials} & \multicolumn{1}{c}{Thermal limit} \\
    \hline\hline\\
   $1D$ & $\begin{array}{ll}\frac{F^{(1)D}}{L} \to 0 & \varepsilon _F = 0\\\frac{F^{(1)D}}{L} = - \frac{1}{\left[d \sqrt{\ln (\frac{2.2 d}{w})} \right]^3} \sqrt{\frac{g e^2 \hbar v_F}{32 \pi^3}} & \varepsilon _F > 0 \end{array}$ & $\begin{array}{ll} \frac{F^{(1)S}}{L} = -  \frac{\mathcal{A}_{1S}g^2 e^4 \hbar^2}{m \Delta^2 d^6}  & \frac{m \varepsilon _F^4 d^4}{g^2 e^4 \hbar^2 \Delta} \ll 1  \\  \frac{F^{(1)S}}{L} = - \frac{1}{\left[d \sqrt{\ln (\frac{2.2 d}{w})} \right]^3} \sqrt{\frac{g e^2}{32 \pi^3} \sqrt{\frac{2 \varepsilon _F \hbar^2}{m}}} & \frac{m \varepsilon _F^4 d^4}{g^2 e^4 \hbar^2 \Delta} \gg 1 
   \end{array} $ & $\frac{F^{(1)T}}{L} = - \frac{\pi k_B T}{8[d \ln (\frac{6.5 d}{w})]^2}$ \\
    \hline
       $2D$ & $\begin{array}{ll}\frac{F^{(2)D}}{A} = - \frac{\mathcal{A}_{2D} g e^2}{d^4}  & \frac{\varepsilon _F d}{g e^2} \ll 1 \\  \frac{F^{2(D)}}{A} = - \frac{\mathcal{A}_{2S} \sqrt{g e^2 \varepsilon _F / 2}}{d^{7/2}}  & \frac{\varepsilon _F d}{g e^2} \gg 1 \end{array}$  & $ \begin{array}{ll} \frac{F^{(2)S}}{A} = - \frac{ (1 -\ln 2) g^2 e^4}{2 \pi \Delta d^5} &  \frac{\varepsilon _F d}{g e^2} \ll 1 \ll \frac{\Delta d}{g e^2} \\  \frac{F^{(2)S}}{A} = - \frac{\mathcal{A}_{2S} \sqrt{g e^2 \varepsilon _F}}{d^{7/2}}  & \frac{\varepsilon _F d}{g e^2} \gg 1 \end{array}$ &  $ \frac{F^{(2)T}}{A}= -\frac{\zeta (3) k_B T}{8 \pi d^3} $ \\
   \hline
   $3D$ & $\frac{F^{(3)D}}{A} = - \frac{\mathcal{A}_{3D} \varepsilon _{max}}{8\pi^2 d^3} \; \; \; 0 \leq \varepsilon _F \ll \varepsilon _{max} $ & $ \begin{array}{ll} \frac{F^{(3)S}}{A} = - \frac{\mathcal{A}_{3S}}{8\pi^2 d^3} \left( \frac{g^2 e^4 m}{\hbar^2} \right)^{1/3} \Delta ^{2/3} & \Delta > 0\\
   \frac{F^{(3)S}}{A} = - \frac{\mathcal{A}_{3S}^{\prime}}{8\pi^2 d^3} \left( \frac{g^2 e^4 m}{\hbar^2} \right)^{1/4} \varepsilon_F ^{3/4}  & \Delta = 0
\end{array}   
   $ & $ \frac{F^{(3)T}}{A} = - \frac{\zeta(3) k_B T}{8 \pi d^3} $\\
   \hline\hline\
    \end{tabular}
    \
}
\end{table*}

\begin{figure}[ht]
    \centering
    \includegraphics[width = 0.45 \textwidth]{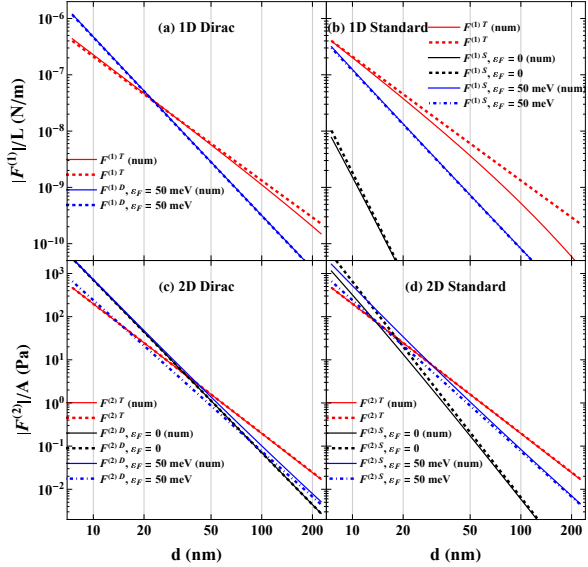}
    \caption{The vdW force per unit length ($L$) or unit area ($A$) in the quantum and thermal limits for (a) $1$D Dirac materials, (b) $1$D standard materials, (c) $2$D Dirac materials and (b) $2$D standard materials.  Here $T = 300 \; \rm{ K}$, $\tau = 0.13 \; \rm{ ps}$, $\varepsilon _F = 50 \; \rm{ meV}$, $g = 4$, $v_F = c/300$ for Dirac materials and $g = 2$, $m = m_e$, $\Delta = 1 \; \rm{ eV}$ for standard materials. The abbreviation (num) denotes calculations obtained by numerical evaluations of all quantities using Eqs. \Eref{eqn:general}, \Eref{eqn:density} and \Eref{eqn:chi}, while the rest of the data are found by the asymptotic scaling laws given in Table \ref{tab:3}.}
    \label{fig:3}
\end{figure}

The scaling laws in the vdW force at 2D are quite diverse. Dirac layers whose Fermi levels pass through the Dirac points interact with a force $F^{(2)D}\sim{d^{-4}}$. We note that a similar expression $F^{(2)D}=-\frac{3g e^2}{128 \pi d^4}$ was obtained for two graphene layers in the retarded Casimir regime, where the Lifshitz formalsim with accurate calculations of the optical response were used \cite{Drosdoff2010, Sarabadani2011, Klimchitskaya2014, Klimchitskaya2020, Lu2021}. Such an expression was also found in other 2D Dirac materials by taking into account Hall conductivity response showing that topologically nontrivial features do not affect significantly the dispersive interaction at closer separations \cite{Rodriguez-Lopez2017, Fialkovsky2018, Lu2021}. Actually, we find that the numerical constant $\mathcal{A}_{2D} = \frac{1}{16 \pi} \int_0^{+\infty} \int_0^{+\infty} \frac{e^{-2u} u^2 du dv}{(1+v/u)^2-e^{-2u}} \approx 0.007641$  compares very well with $\frac{3}{128\pi}$. The difference of about $2\%$ between the two results indicates that retardation and contributions from the transverse modes are not important in the dispersive interaction in such materials.   
 One further notes that this interaction has the same scaling law as in perfect metals ($F_m=-\frac{\hbar c \pi^2}{240 d^4}$), however the magnitude is much reduced (at least by two orders). Notably, the scaling law $d^{-4}$ is consistent with any polarization function whose function dependence is $\chi^{(2)}_0 \sim q f (\omega / q)$ as long as $\varepsilon_F=0$ \cite{Santos2009}. When $\varepsilon_F>0$, the quantum vdW force for Dirac and standard materials displays $d^{-7/2}$ behavior. Such a scaling dependence is consistent with results found by others \cite{Sernelius1998, Dobson2001, Bostrom2000, Dobson2006, Pablo2014} when considering non-retarded dispersive interactions between different metals. 
The presence of an energy gap in the parabolic band structure changes the distance dependence to $F^{(2)D}\sim{d^{-5}}$ indicating a longer ranged interaction when comparing 2D metals and dielectrics  \cite{Rydberg2003, Dobson2006, Stedman2014}.

For 3D systems, $F^{(3)}\sim d^{-3}$ for all cases showing a universal scaling law in the non-retarded regime. It is interesting to note that for Dirac materials, the explicit dependence of the polarization function upon $\varepsilon_{max}$ specifying the validity of linear dispersion carries over to the vdW interaction as seen in the constant $\mathcal{A}_{3D}$ (which in general depends on $\varepsilon_{max}$, see Supplementary Material). In fact, the $F^{(3)}\sim d^{-3}$ scaling functionality is also valid at larger separations when retardation is included and this is the case not only for Dirac materials, but also for Weyl semimetals whose 3D Dirac energy cones are non-degenerate \cite{Rodriguez2020, Bordag2021}. For standard systems, the quantum vdW force directly depends on the Fermi level in the case of metals or the band gap in the case of dielectrics.

Next we consider the thermal limit of the vdW interaction by examining the $l=0$ term in Eq. \Eref{eqn:general}. For this purpose, the polarization function is calculated at finite temperature by utilizing the Maldague formula \cite{Maldague1978, Ando1982},
\begin{eqnarray}\label{eqn:relax-temp}
    \chi^{(N)}_{0} && \left(\mathbf{q}, \imath \omega _l, T > 0, \varepsilon _F, \tau \right) = \nonumber\\
    && = \int _0^{+\infty} \frac{\chi^{(N)}_{0} \left(\mathbf{q}, \imath \left( \omega _l + \frac{1}{2 \tau} \right), T = 0, \varepsilon _F^{\prime} \right)}{4 k_B T \cosh ^2 \left( \frac{ \varepsilon _F - \varepsilon _F^{\prime} }{ 2 k_B T } \right) }  d \varepsilon _F^{\prime},
\end{eqnarray}
which also takes into account the relaxation time $\tau$ (assumed to be constant) due to scattering processes in the materials. The results from our calculations are given in Table \ref{tab:3}. They indicate that the thermal vdW interaction is primarily determined by the dimensionality of the objects and the materials properties play a secondary role. For 1D systems $F^{(1) T}\sim [d \ln (6.5d/w)]^{-2}$, while the scaling law $d^{-3}$ is the same for 2D and 3D interacting materials. 

\begin{figure}[ht]
    \centering
    \includegraphics[width = 0.45 \textwidth]{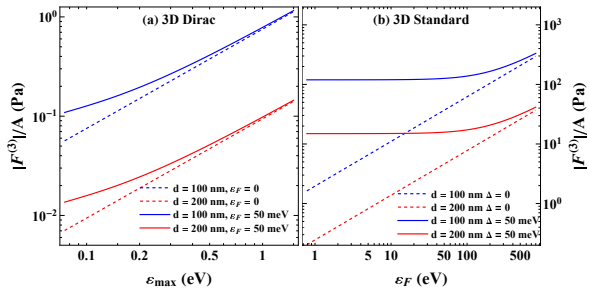}
    \caption{The vdW force per unit area ($A$) in $3$D: (a)  as a function of bandwidth $\varepsilon_{max}$ for Dirac materials and (b) as a function of Fermi level $\varepsilon _F$ for standard materials. Here $T = 300 \; \rm{ K}$, $\tau = 0.13 \; \rm{ ps}$, $\varepsilon _F = 50 \; \rm{ meV}$, $g = 4$, $v_F = c/300$ for Dirac materials and $g = 2$, $m = m_e$ for standard materials.}
    \label{fig:4}
\end{figure}

To gain further insight into the vdW interaction, the quantum mechanical and thermal forces are shown graphically as a function of distance in Fig. \ref{fig:3}. For 1D Dirac materials (Fig. \ref{fig:3}a), we find that  the two limits coincide at $d_T \approx 20 \; \rm{ nm}$ and at larger separations the magnitude of $F^{(1) T}$ becomes larger due to the slower distance decay compared to $F^{(1) D}$. For standard 1D interacting materials, the quantum vdW attraction is always smaller than the thermal limit in the range of interest $10-200 \; \rm{ nm}$. In the case of 1D materials for which $\Delta \neq 0$, $F^{(1) D}$ falls very fast as the separation is increased, as can be seen in Fig. \ref{fig:3}b indicating the dominance of thermal fluctuations. 

For 2D materials, cross-over between different regimes is also observed. Particularly for $2$D Dirac materials with $\varepsilon_F=0$, the thermal force begins to dominate over the quantum limit for $d > d_T \approx 35 \; \rm{ nm}$, while this happens at $d > d_T \approx 40 \; \rm{ nm}$ when $\varepsilon=50$ meV for the two Dirac layers (Fig. \ref{fig:3}c). Transitioning from a quantum mechanical to a thermal limit is also observed in Fig. \ref{fig:3}d for standard materials, although the distance where that occurs is reduced. For example, the thermal force becomes larger at $d > d_T \approx 15 \; \rm{ nm}$ for parabollic materials with $\varepsilon_F=0$, while this effect is found for $d_T \approx 30 \; \rm{ nm}$ for parabollic materials with $\varepsilon _F = 50 \; \rm{ meV}$.

For 3D materials, the vdW force is always $\sim d^{-3}$. For Dirac systems, the quantum force depends linearly on the bandwidth $\varepsilon_{max}$ as shown in Fig. \ref{fig:4}a. For standard materials, the interaction can be modulated by the Fermi level, as shown in Fig. \ref{fig:4}b. As $\varepsilon_F$ lies inside a non-zero energy gap, the quantum vdW is completely determined by $\Delta$ at a given distance according to $F^{(3)S}\sim\Delta^{2/3}$. When $\Delta=0$ or $\varepsilon_F$ is in the conduction or valence region, $F^{(3)S} \sim \varepsilon_F^{3/4}$ as shown in Fig. \ref{fig:4}b. 

\begin{figure}[ht]
    \centering
    \includegraphics[width = 0.45 \textwidth]{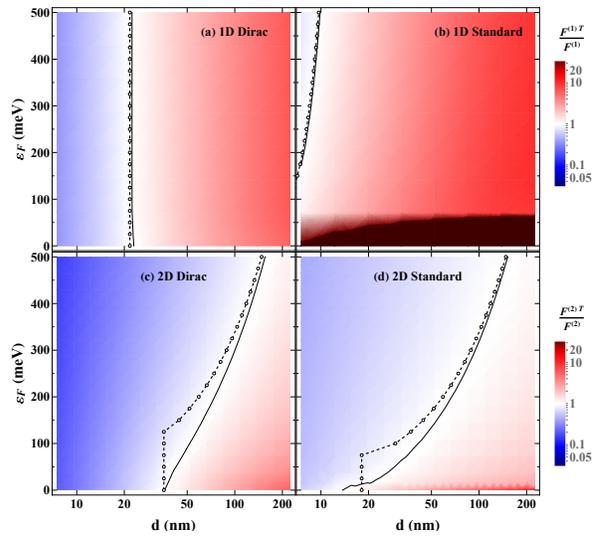}
    \caption{Density plots of the ratio $F^{(N)D,S}/F^{(N) T}$ between the quantum and thermal vdW force at $T = 300 \; \rm{ K}$ in $\varepsilon _F$ - $d$ space. {The solid black lines obtained numerically by using Equations (S-19), (S-21) and (S-23) in the Supplementary Material, and the dot-dashed lines obtained analytically by using the asymptotic scaling laws in Table \ref{tab:3}}, show when the limits {coincide}. Regions in blue correspond to the dominance of quantum mechanical effects, while regions in red correspond to dominating thermal effects, according to the given color bar.  Here $\tau = 0.13 \; \rm{ ps}$, $g = 4$, $v_F = c/300$ for Dirac materials and $g=2$, $m = m_e$, $\Delta = 1 \; \rm{ eV}$ for standard materials.}
    \label{fig:5}
\end{figure}

A question of fundamental importance is how to better understand the characteristic distance $d_T$  separating the quantum mechanical and thermal limits of the nonretarded vdW interaction involving different materials. Given the analytical expressions in Table \ref{tab:3}, equating  $F^{(N)D,S}=F^{(N)T}$ gives the means of finding $d_T$ at which the quantum-to-thermal transition occurs. In Fig. \ref{fig:5}, we show contour plots of $\frac{F^{(N)D,S}}{F^{(N)T}}$ in the $\varepsilon_F$ vs $d$ map, where $d_T$ is clearly marked. It is interesting to note that in 1D materials (Fig. \ref{fig:5}a,b), the thermal effects are especially strong. In systems with linear band dispersion, we find that $d_T = \frac{\sqrt{2 g e^2 \hbar v_F}}{\pi^{5/2} k_B T} \ln ^2 \left( \frac{6.5 \sqrt{2 g e^2 \hbar v_F}}{\pi^{5/2} w k_B T} \right) \ln ^{-3/2} \left( \frac{2.2 \sqrt{2 g e^2 \hbar v_F}}{\pi^{5/2} w k_B T} \right)$, which does not depend on the Fermi level $\varepsilon _F$.  In standard materials, however, practically the entire region is dominated by thermal fluctuations. Separations smaller than $10 \; \rm{ nm}$  must be considered (which may involve taking into account the atomic structure of the materials) in order to have dominating quantum effects in the vdW interaction for 1D materials.

Significant thermal effects are also found in 2D materials, as shown in Fig. \ref{fig:5}c,d. We obtain that for small $\varepsilon_F$ where $F^{(2)D}\sim 1/d^{-4}$, $d_T = \frac{8 \pi A_{2D} g e^2}{\zeta (3) k_B T}$. For larger $\varepsilon_F$ where $F^{(2)D}\sim 1/d^{-7/2}$, $d_T = \left(\frac{8 \pi A_{2D}}{\zeta (3)}\right)^2 \frac{g e^2 \varepsilon_F}{k_B^2 T^2}$ . These results show that in the first situation,  $d_T$ is independent of the Fermi level and it is $\sim T^{-1}$, while in the second situation $d_T \sim \varepsilon_F/(k_BT)^2$. In the case of 2D standard materials, for $F^{(2)S}\sim 1/d^{-5}$ corresponding to small Fermi levels,  the characteristic thermal distance $d_T = \sqrt{\frac{4 (1-\ln 2)}{\zeta(3)}} \frac{g e^2}{\sqrt{k_B T \Delta}} \sim T^{-1/2}$, while for $F^{(2)S}\sim 1/d^{-7/2}$ corresponding to large Fermi levels, $d_T = \left(\frac{8 \pi A_{2D}}{\zeta (3)}\right)^2 \frac{g e^2 \varepsilon_F}{k_B^2 T^2} \sim \varepsilon _F / (k_BT)^2$. 

For 3D materials, the quantum-to-thermal transition is controlled by other properties entering the expressions for the interaction, since the $d^{-3}$ scaling law is the same for both limits. We find that $\frac{F^{(3) T}}{F^{(3) D}} \sim \frac{k_B T}{\varepsilon_{max}}$, which emphasizes again the importance of the bandwidth specifying the validity of the linear dispersion. For 3D parabolic materials,  $\frac{F^{3T}}{F^{3S}} \sim \left(\frac{\hbar^2}{g^2 e^4 m}\right)^{1/4} \frac{k_B T}{\varepsilon_{F}^{3/4}}$ or $\frac{F^{(3)T}}{F^{(3)S}} \sim \left(\frac{\hbar^2}{g^2 e^4 m}\right)^{1/3} \frac{k_B T}{\Delta^{2/3}}$, which shows that the onset of thermal fluctuations can be controlled via the Fermi level or nonzero band gap in a similar way. To get an idea about the thermal characteristic distance in the non-retarded vdW regime, one must go beyond the long wavelength approximation and perhaps include retardation effects. This, however, may involve a different method of calculations, which goes beyond the scope of this paper.

\section{Conclusions}

In summary, the dispersive vdW interaction has been studied with the RPA method for materials with parabolic and linear energy dispersions. The quantum mechanical and thermal regimes are investigated in the case of interacting identical 1D, 2D, and 3D systems. We find that there is an intricate and potentially tunable by the Fermi level relationship between dimensionality and response properties relationship in the dispersive force. 

The interaction between 1D materials is dominated by thermal fluctuations regardless of the energy band structure. In 2D materials, however, the characteristic distance for the quantum-to-thermal transitions depends strongly on the Fermi level with the general trend that larger $\varepsilon_F$ results in larger $d_T$. By changing $\varepsilon_F=(0,200) \; \rm{ meV}$, $d_T \sim (35,75) \; \rm{ nm}$ for Dirac and $d_T \sim (15,70) \; \rm{ nm}$ for standard materials. Similar results for the onset of thermal effects can be found for the retarded Casimir interaction between undoped graphene layers for which $d_T=\hbar v_F/k_B T \sim 30 \; \rm{ nm}$ \cite{Khusnutdinov2018,Drosdoff2012,Liu2021,Bimonte2017,Klimchitskaya2015}. Our calculations show that such characteristic thermal distances are typical not only for Dirac materials, but also for standard parabolic systems in 2D. Comparing with the Casimir thermal distance for 3D metals $d_T=\hbar c/k_B T = 7.6\times 10^3 \; \rm{ nm}$ shows that thermal fluctuations become prominent at much smaller separations. It further appears that dimensionality is a leading factor in determining the onset of thermal effects, while the materials properties may be less important with the exception of 2D Dirac materials with $\varepsilon_F=0$. In this case, regardless of  $\varepsilon_F$, $d_T$ for the non-retarded regime is always smaller by at least an order comparing with the micrometer $d_T$ range  that is typical for the retarded Casimir regime for 3D materials.  

It is also interesting to note that the onset of thermal effects in the vdW regime in 3D is controlled by the ratio of $k_BT$ and various energy parameters ($\varepsilon_{max}, \varepsilon_F, \Delta$) of the materials, as discussed above. In all cases however, the thermal contribution becomes more prominent with temperature, but this effect can be modulated by other band structure properties. In the case of 3D Dirac materials, for example,  $F^{3T}>F^{3D}$ occurs for $\varepsilon_{max}<0.023$ eV at $T=300$ K, since $\varepsilon_{max}<k_BT$. The explicit dependence upon $\varepsilon_{max}$ is a direct consequence of the presence of the bandwidth in the polarization function. Since the scaling law for the interaction in 3D is the same regardless of the energy bands, we conclude that the onset of thermal effects can occur at any separation providing $k_BT$ is large enough or $\varepsilon_{max}, \varepsilon_F, \Delta$ parameters are small enough to ensure $F^{3T}>F^{3D}$.

This investigation gives a comprehensive understanding of the characteristic functionalities and asymptotics of vdW interactions at separations where retardation can be neglected. Our studies highlight the interplay between energy dispersion and dimensionality in the quantum mechanical and thermal limits of the vdW force. Interestingly, we find that thermal fluctuations for systems with reduced dimensions can become pronounced at much smaller separations when compared to their 3D counterparts. Our results provide a comprehensive picture of nonretarded dispersive interactions, which is complementary to previous in-depth studies focusing on the retarded Casimir regime \cite{Woods2016, Klimchitskaya2009}. They can serve as a useful guidance to future experiments that might be geared towards demonstrating the pronounced role of thermal fluctuations at much reduced separations, a previously unexplored area.

\section{Acknowledgments}
L.M.W. acknowledges financial support from the US Department of Energy under grant No. DE-FG02-06ER46297. {P. R.-L. was supported by "AYUDA PUENTE 2021, URJC"}.

\bibliographystyle{iopart-num}
\bibliography{ref}

\newpage
\thispagestyle{empty}

\begin{figure*}
    \centering
    \includegraphics[width = \linewidth]{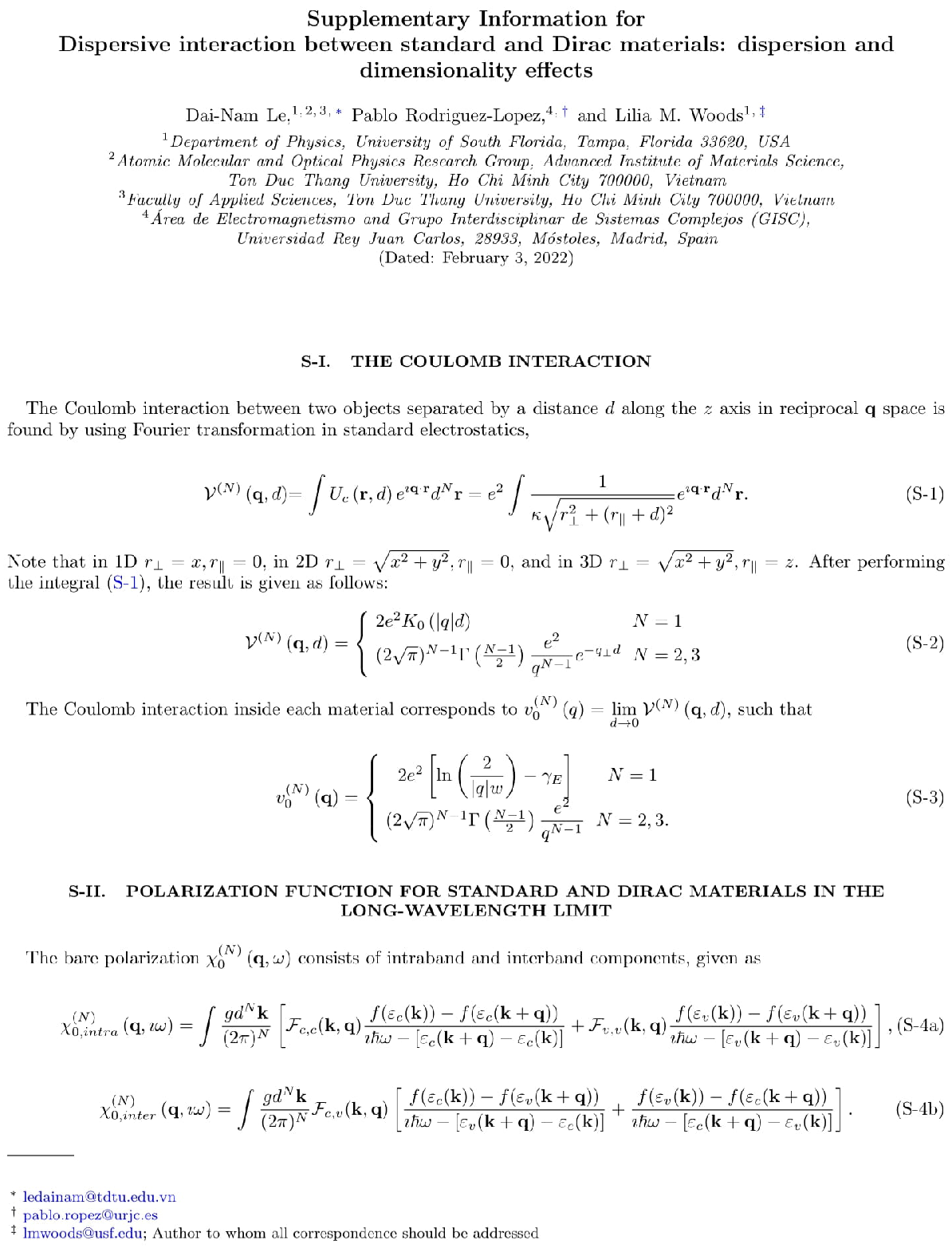}
\end{figure*}

\thispagestyle{empty}
\begin{figure*}
    \centering
    \includegraphics[width = \linewidth]{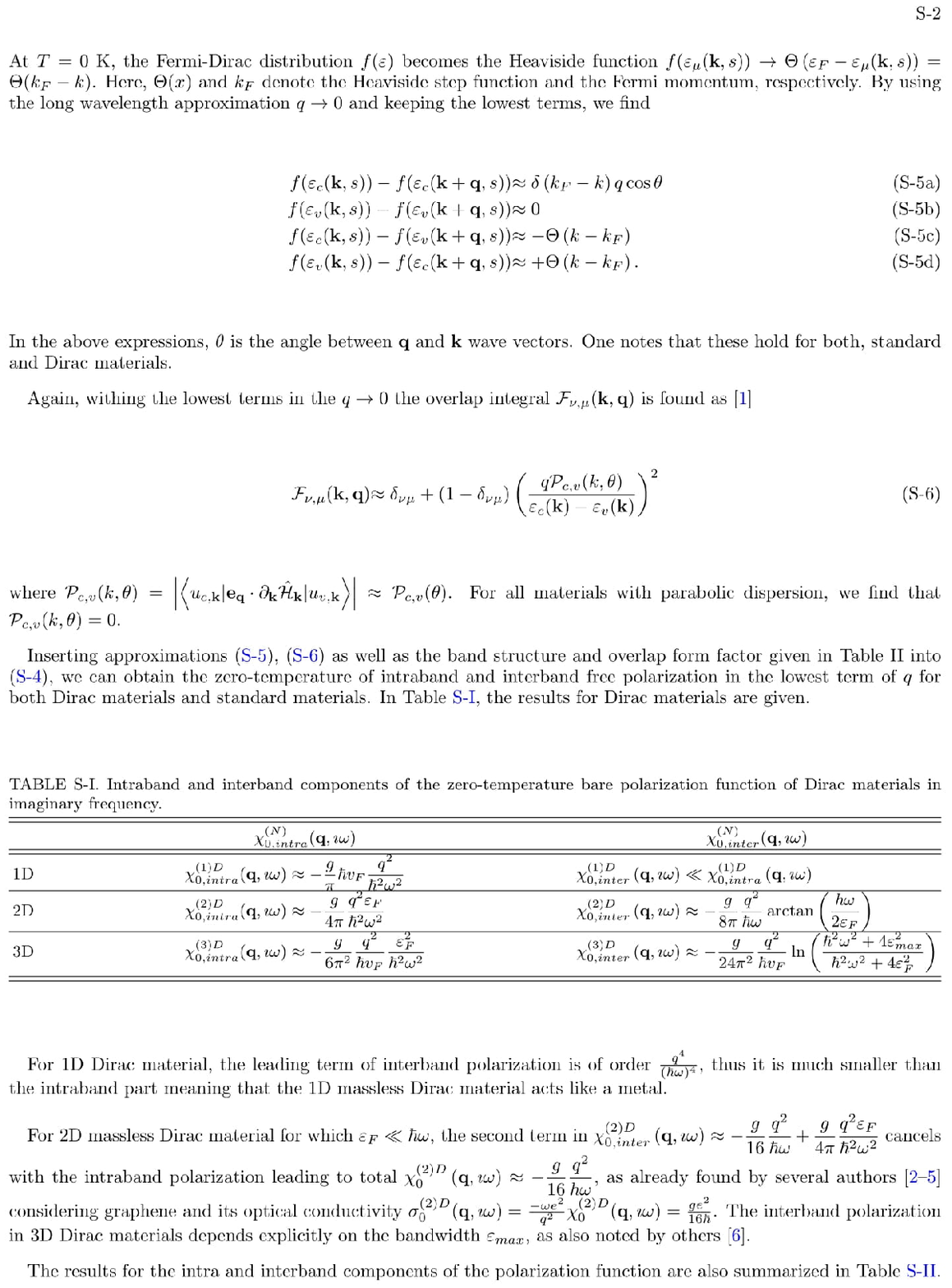}
\end{figure*}

\thispagestyle{empty}
\begin{figure*}
    \centering
    \includegraphics[width = \linewidth]{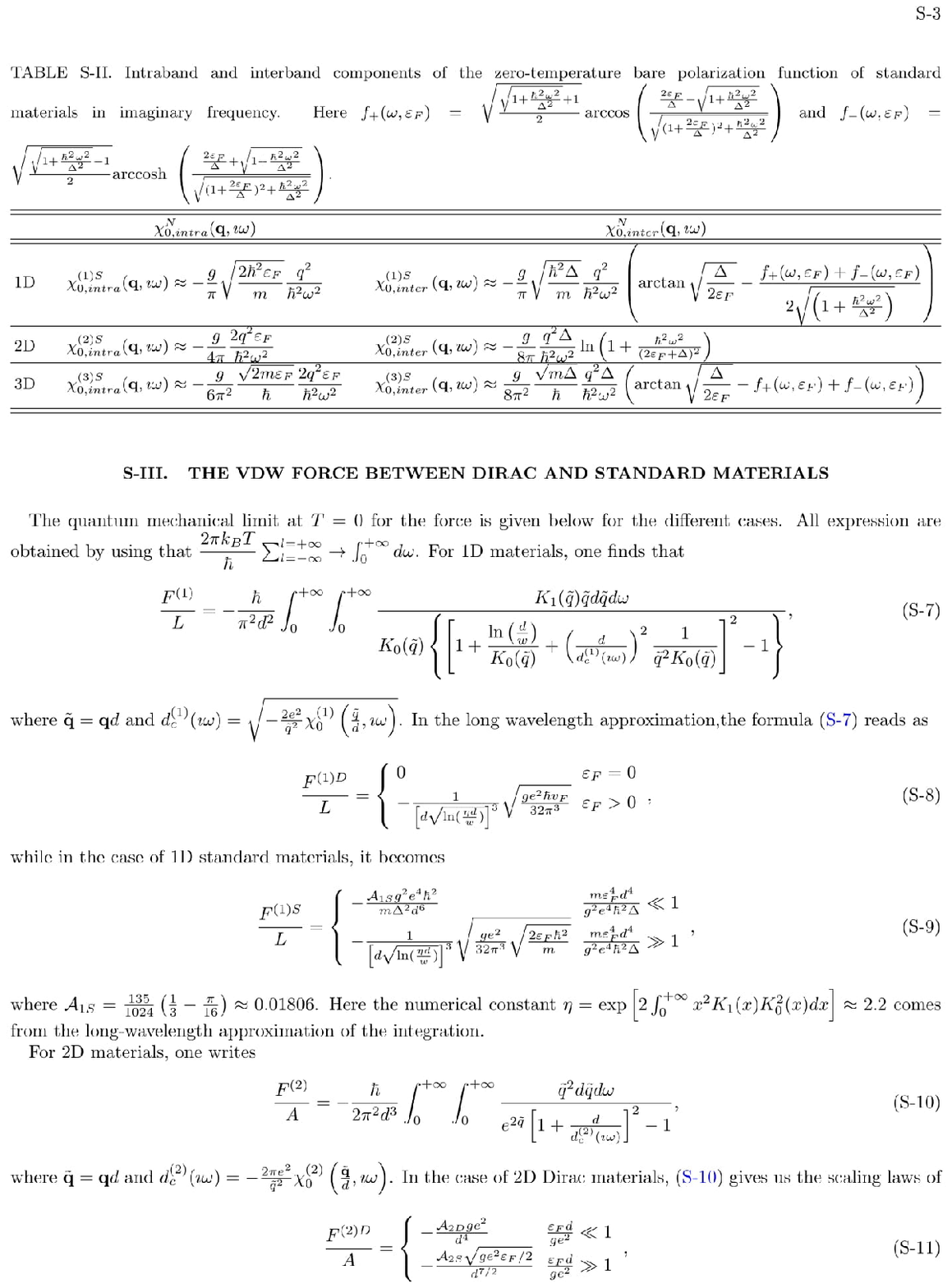}
\end{figure*}

\thispagestyle{empty}
\begin{figure*}
    \centering
    \includegraphics[width = \linewidth]{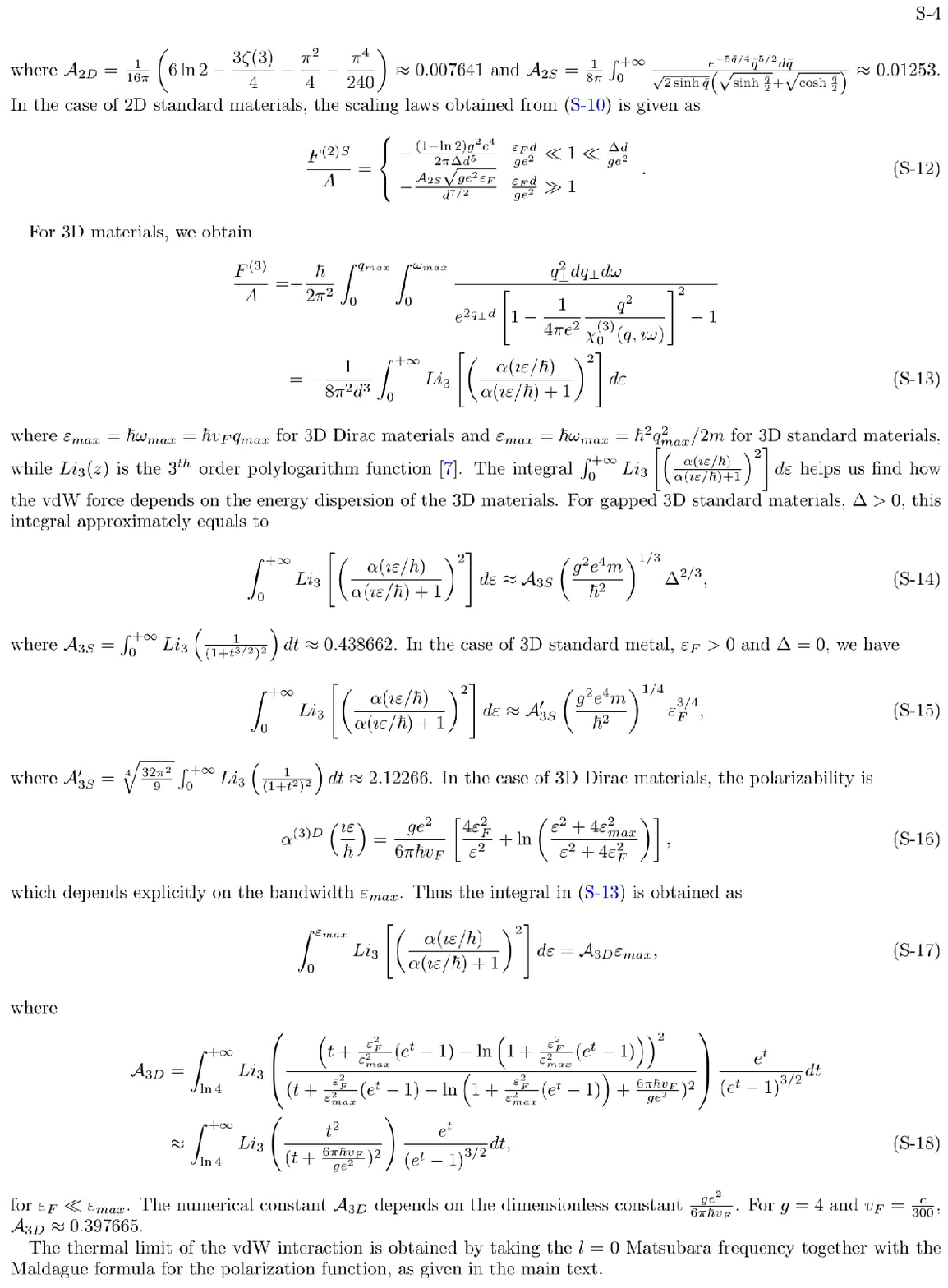}
\end{figure*}

\thispagestyle{empty}
\begin{figure*}
    \centering
    \includegraphics[width = \linewidth]{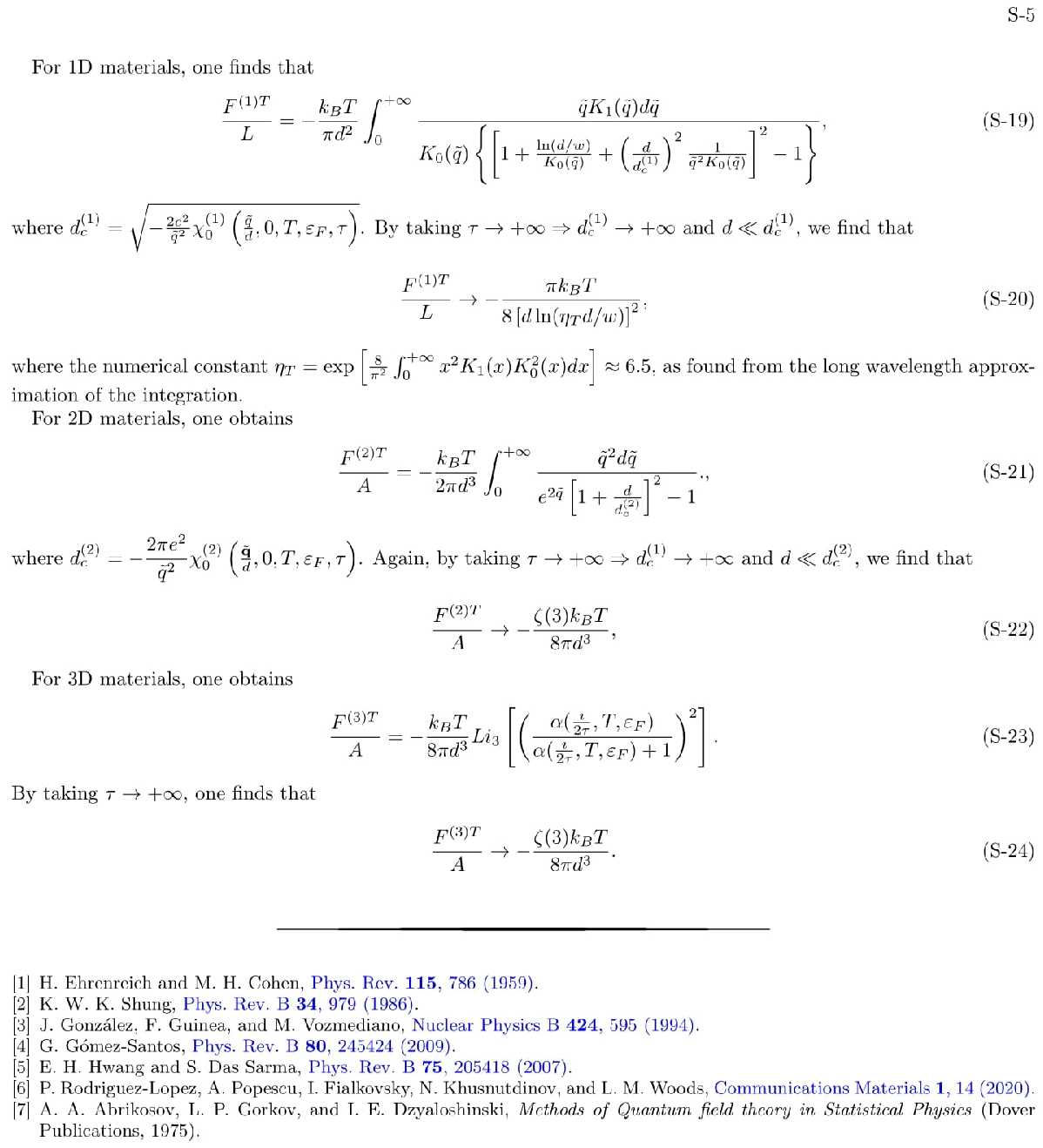}
\end{figure*}
\end{document}